\documentclass[twocolumn,showpacs,preprintnumbers,amsmath,amssymb,floatfix]{revtex4}

\usepackage{graphicx}
\usepackage{dcolumn}
\usepackage{bm}

\newcommand{\beq}{\begin{equation}}
\newcommand{\eeq}{\end{equation}}
\newcommand{\bey}{\begin{eqnarray}}
\newcommand{\eey}{\end{eqnarray}}

\begin{document}

\preprint{}

\title{A class of interior solutions corresponding to a $(2+1)$ dimensional asymptotically anti-de Sitter spacetime}

\author{Ranjan Sharma}
 \email{rsharma@iucaa.ernet.in}
\affiliation{Department of Physics, P. D. Women's College, Jalpaiguri 735101, India.}

\author{Farook Rahaman}
\email{farook\_rahaman@yahoo.com} \affiliation{Department of
Mathematics, Jadavpur University, Kolkata 700 032, West Bengal,
India.}

\author{Indrani Karar}
\email{indrani.karar08@gmail.com} \affiliation{Department of
Mathematics, Saroj Mohan Institute of Technology, Guptipara, West
Bengal, India.}

\date{\today}

\begin{abstract}
Lower dimensional gravity has the potential of providing non-trivial and valuable insight
into some of the conceptual  issues arising in four dimensional relativistic gravitational
analysis. The asymptotically anti-de Sitter ($2+1$) dimensional spacetime described by Ba$\tilde{n}$ados,
Teitelboim and Zanelli (BTZ) which admits a black hole solution, has become a source of fascination for
theoretical physicists in recent years. By suitably choosing the form of the mass function $m(r)$,
 we derive a new class of solutions for the interior of an isotropic star corresponding to the exterior
  $(2+1)$ asymptotically anti-de Sitter BTZ spacetime. The solution obtained satisfies all the regularity
  conditions and its simple analytical form helps us to study the physical parameters of the configuration
  in a simple manner.
\end{abstract}


\maketitle

\section{Introduction}
It is well known that lower dimensional gravity has the potential of providing
non-trivial and valuable insight into some of the conceptual  issues arising in
 four dimensional relativistic gravitational analyses. For example, the ($2+1$)
 dimensional spacetime geometry described by Ba$\tilde{n}$ados, Teitelboim and
 Zanelli\cite{BTZ} (henceforth BTZ), which is asymptotically anti-de Sitter and
 admits a black hole solution, has become a source of fascination for theoretical
 physicists in recent years. The advantage here is that though the system mimics
 $4$-dimensional analysis, it offers less intricate set of equations to deal with.
 It is, therefore, crucial to find a physically reasonable interior solution
  corresponding to the exterior BTZ spacetime which may give us deep understanding
  about the nature of a gravitationally collapsing body and its consequences. Mann
  and Ross\cite{Mann} analyzed the collapse of a $(2+1)$ dimensional star filled with dust
  ($p=0$) corresponding to the exterior BTZ spacetime and showed under what condition it
  might collapse to a black hole.  For an incompressible fluid ($\rho =constant$), the
  interior solution obtained by Cruz and Zanelli\cite{Cruz} puts a bound on the maximum
  allowed mass for such an object. The study also claims that the collapsed stage would
   always be covered under its event horizon. Cruz {\em et al}\cite{Cruz2} presented a
   new solution corresponding to an exteripr BTZ  spacetime with the choice $\rho = \rho_c(1-r^2/R^2)$,
   where, $\rho_c$ is the central density, $\rho$ is the density which is a function of
   the radial parameter $r$, and $R$ is the boundary of the star. Another solution has been
   reported by Paulo M. S$\acute{a}$\cite{Paulo} where a polytropic equation of state (EOS)
    of the form $p=K\rho^{1+1/n}$ was assumed, where $n$ is the polytropic index and $K$ is
     the polytropic constant.

In this paper, we report a  new class of solution corresponding to
the BTZ exterior spacetime which has been obtained by assuming a
particular form of the mass function $m(r)$. The solution obtained
has been found to satisfy all the regularity conditions and its
simple analytical form helps us to study the physical parameters
of the configuration in a simple manner.

\section{Interior spacetime}
The metric for the interior of a static spherically symmetric distribution of matter in $(2+1)$
dimensions has the standard form
\begin{equation}
ds^2 = -e^{2\gamma(r)} dt^2 + e^{2\mu(r)} dr^2 + r^2d\theta^2. \label{eq1}
\end{equation}
We assume that the energy-momentum tensor for the matter distribution at the interior of the star is given by
\begin{equation}
T_{ij} = (\rho + p ) u_i u_j + p  g_{ij}, \label{eq2}
\end{equation}
where, $\rho$ represents the energy density, $p$ is  the
isotropic pressure, and  $u^{i}$ is the $3$-velocity of the
fluid. The Einstein's field equations with a negative
cosmological constant ($\Lambda < 0$), for the spacetime given in
Eq.~(\ref{eq1}) together with the energy-momentum tensor given in
Eq.~(\ref{eq2}), rendering
 $G = c = 1$, yield the following independent equations
\begin{eqnarray}
2\pi \rho +\Lambda &=& \frac{\mu' e^{-2\mu}}{r}, \label{eq3} \\
2\pi p  -\Lambda &=& \frac{\gamma' e^{-2\mu}}{r}, \label{eq4}  \\
2\pi p  -\Lambda &=&
e^{-2\mu}\left(\gamma'^2+\gamma''-\gamma'\mu'\right),\label{eq5}
\end{eqnarray}
where a `$\prime$' denotes differentiation with respect to the radial parameter $r$.
 Combining Eqs.~(\ref{eq3})-(\ref{eq5}), we have
\begin{equation}
\left(\rho + p\right)\gamma' + p '   =0,\label{eq6}
\end{equation}
which is the conservation equation in $(2+1)$ dimensions. Since the system has three
independent equations involving four unknown parameters ($\rho$, $p $, $\gamma(r)$,
 and $\mu(r)$), we will always have one degree of freedom to solve the set of equations.
 In some previous works, this freedom has been utilized to model (i) a dust cloud ($p=0$)
 \cite{Mann}; (ii) a fluid of uniform density ($\rho=constant$)\cite{Cruz}; and (iii)
  a polytropic EOS $p=K\rho^{1+1/n}$\cite{Paulo}. In this paper, we choose a physically
   meaningful form of the mass function $m(r)$, expressible in terms of the metric function $\mu(r)$ in the form
\begin{equation}
2m(r) = C - e^{-2\mu(r)} - \Lambda r^2,\label{eq7}
\end{equation}
which can be obtained by integrating Eq.~(\ref{eq3}). In Eq.~(\ref{eq7}), $C$ is an
integration constant and $m(r)$ is the mass within a  radial distance $r$ defined as
\begin{equation}
m(r) = \int_0^r 2\pi \rho \tilde{r}d\tilde{r}.\label{eq8}
\end{equation}
We assume that $m(r)$ has the form
\begin{equation}
m(r) = \frac{1}{2}-
\frac{\Lambda}{2} r^2 - \frac{e^{-Ar^2}}{2}, \label{eq9}
\end{equation}
which is regular at the centre, i.e., $m(r)=0$ at $r=0$, $A$ being a constant. This implies
\begin{equation}
2\mu(r)= A r^2,\label{eq10}
\end{equation}
where, without any loss of generality, we have set $C=1$. Note that, for $g_{rr} > 0$, we must have
\begin{equation}
m(r) < \frac{1}{2} - \frac{\Lambda r^2}{2}.\label{eq11}
\end{equation}
From Eq.~(\ref{eq3}), the energy density is then obtained as
\begin{equation}
\rho = \frac{1}{2\pi}\left[Ae^{-Ar^2} - \Lambda\right].\label{eq12}
\end{equation}
Combining Eqs.~(\ref{eq4}), (\ref{eq5}) and (\ref{eq10}), we get,
\begin{equation}
\gamma'^2+\gamma''-Ar\gamma' = \frac{\gamma'}{r}, \label{eq13}
\end{equation}
which on integration yields,
\begin{equation}
\gamma'  = \frac{Ar  e^{\frac{A}{2}r^2}}{BA +
e^{\frac{A}{2}r^2}}, \label{eq14}
\end{equation}
where $B$ is an integration constant. Further integrating Eq.~(\ref{eq14}) we get
\begin{equation}
\gamma  = \ln [ D( BA + e^{\frac{A}{2}r^2})],\label{eq15}
\end{equation}
where $D$ is another integration constant. The spacetime metric
thus obtained is free from any central singularity problem. The
isotropic pressure is then obtained as
\begin{equation}
p  = \frac{\frac{A}{2\pi}  e^{-\frac{A}{2}r^2}}{BA +
e^{\frac{A}{2}r^2}} + \frac{\Lambda}{2\pi}\label{eq16}
\end{equation}

Before we analyze physical features of the model, let us write $\Lambda = -K^2$ to
ensure that the cosmological constant always remains negative. The central density
and pressure are then obtained as
\begin{eqnarray}
\rho_c = \rho(r=0) &=& \frac{1}{2\pi}(A  + K^2), \label{eq17}\\
p  (r=0)  &=&  \frac{1}{2\pi}\left(\frac{A }{BA + 1}
 - K^2\right), ~~~~~~~~~\label{eq18}
\end{eqnarray}
Thus, the central density and pressure remain positive if
$\frac{A}{(BA + 1)} > K^2$. For a physically meaningful solution, the radial and tangential
pressures should be
decreasing functions of $r$. In our model, we have
\begin{equation}
\frac{d\rho}{dr} =  -\frac{1}{\pi}r A^2
e^{-Ar^2},\label{eq19}\end{equation}
\[\frac{d p }{dr} = -\frac{1}{2\pi}
\frac{A^2r}{\left(BA+e^{\frac{A}{2}r^2}\right)}\left[
e^{-\frac{A}{2}r^2}+ \frac{1}{\left(BA+e^{\frac{A}{2}r^2}\right)}
\right].\label{eq20}\]
\begin{equation}\label{eq20}\end{equation}
Obviously, at $r=0$, $\frac{d\rho}{dr} = \frac{d p}{dr}  = 0$, and the density and pressure
 decrease radially outward as can be seen in Fig.~(\ref{fig:1}) and (\ref{fig:2}), respectively.

The radius of the star $R$  can be obtained by letting $p(r=R) = 0$ in Eq.~(\ref{eq16}), which gives
\begin{equation}
R = \sqrt{\frac{2}{A}\ln\left[ \frac{1}{2}\left ( -BA+\sqrt{B^2A^2
+\frac{4A}{K^2}}\right)\right]}.\label{eq21}
\end{equation}
The total mass confined within  the radius $R$ is obtained as
\begin{equation}
M_{Total} =  \frac{1}{2} + \frac{K^2 R^2}{2} -
\frac{e^{-AR^2}}{2}.\label{eq22}
\end{equation}

\section{Exterior spacetime and matching conditions}
We match the interior solution to the exterior BTZ metric\cite{BTZ} given by
\begin{equation}
ds_{+}^2 = - \left(- M_0 - \Lambda r^2\right) dt^2 +\left(- M_0 - \Lambda r^2\right)^{-1}dr^2 + r^2 d\theta^2,
\label{eq23}
\end{equation}

where, the parameter $M_0$ is the conserved charge associated
with asymptotic invariance under time displacements. Continuity
of $g_{tt}$, $g_{rr}$ and $\frac{\partial g_{tt}}{\partial r}$
across the circular disc joining the two matrices at $r= R$,
implies
\begin{equation}
\left(- M_0 - \Lambda R^2\right)  =  D^2 \left(
AB+e^{\frac{A}{2}R^2}\right)^2,\label{eq24}\end{equation}
\begin{equation}\left(- M_0 - \Lambda R^2\right)  =  e^{-AR^2},\label{eq25}\end{equation}
\begin{equation}-2  \Lambda R    = 2D^2 \left(
AB+e^{\frac{A}{2}R^2}\right)e^{\frac{A}{2}R^2} AR.
\label{eq26}\end{equation}

Solving the above set of equations simultaneously, we get
\begin{eqnarray}
A &=& - \frac{1}{R^2}\ln\left(K^2 R^2 - M_0\right), \label{eq28}\\
D &=& \frac{-K^2R^2}{\ln\left(K^2 R^2 - M_0\right)},\label{eq29}\\
B &=&  \frac{\left[(K^2R^2-M_0) + \frac{K^2R^2}{\ln\left(K^2 R^2 -
M_0\right)}\right]}{K^2\sqrt{\left(K^2 R^2 - M_0\right)}}.\label{eq30}
\end{eqnarray}

We demand that the following energy conditions should be satisfied by the distribution:
 (i) $\rho,~ p \geq 0$, (ii)
$\rho+p\geq 0$ and (iii) $\rho+2p \geq 0$. Employing these energy conditions at the centre
($r=0$), we note that the first
condition will be satisfied if $\frac{A }{(BA + 1)} > K^2$, the second condition will be
satisfied if $A \geq 0$, and the third
condition demands that $A+\frac{2A}{(BA + 1)} > K^2$.

\begin{figure}
\vspace{0.2cm} \includegraphics[width=0.4\textwidth]{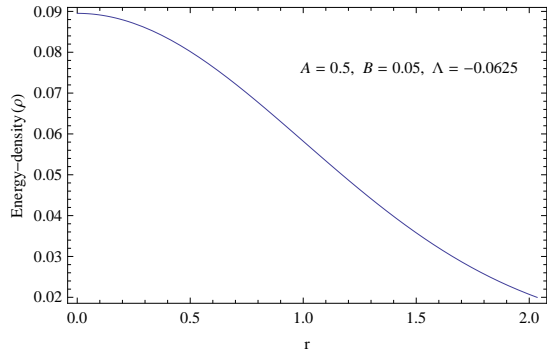}
\caption{Density variation at the stellar interior.} \label{fig:1}
\end{figure}

\begin{figure}
\vspace{0.2cm} \includegraphics[width=0.4\textwidth]{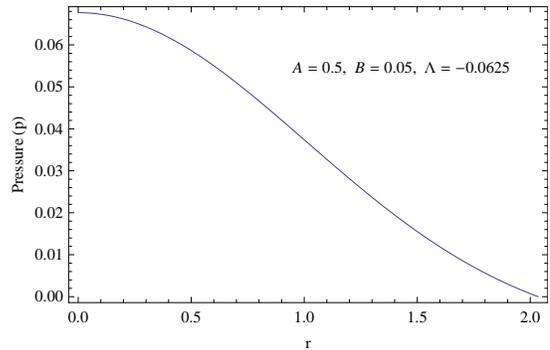}
\caption{Pressure variation at the stellar interior.}
\label{fig:2}
\end{figure}

\begin{figure}
\vspace{0.2cm} \includegraphics[width=0.4\textwidth]{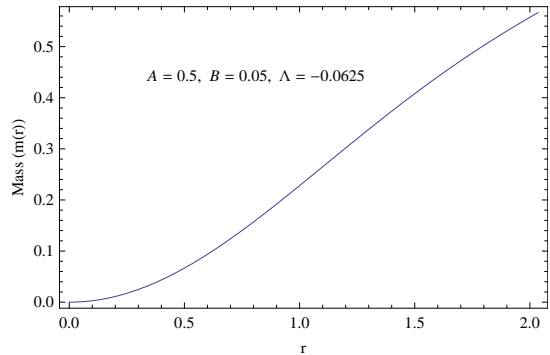}
\caption{Variation of mass function at stellar interior.}
\label{fig:3}
\end{figure}

To understand behaviour of the physical parameters in our model, we assume $A=0.5$,
 $B=0.05$, $\Lambda = -0.0625$ satisfying the above conditions. Solving Eqs.~(\ref{eq28}) and
 (\ref{eq30}), we then get $R= 2.035$, $M_0 = 0.1327$.
 With these values, variations of the physical parameters like energy density, pressure and
 mass function have been shown in Fig.~(\ref{fig:1}), (\ref{fig:2}) and (\ref{fig:3}), respectively.

\section{Some features}

\subsection{Maximum mass-radius relation }
\begin{figure}
\begin{center}
\vspace{0.2cm} \includegraphics[width=0.4\textwidth]{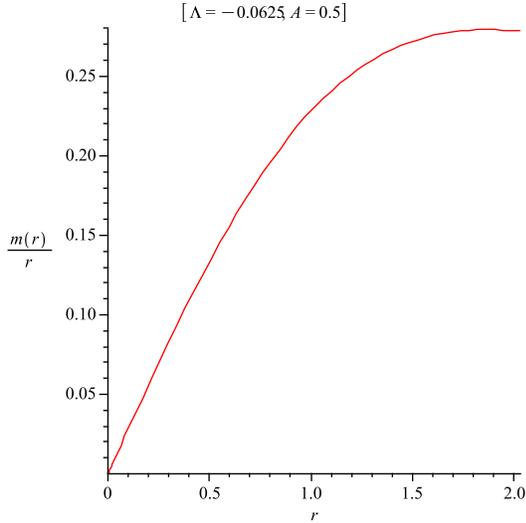}
\end{center}
\caption{The variation of $\frac{m(r)}{r}$  is shown against r.}
\label{fig:4}
\end{figure}

The upper bound on the mass in our model can be obtained from the observation that
\begin{equation}
2m(r) \equiv 1+K^2 r^2- e^{-Ar^2} \leq 1+K^2r^2 -e^{-AR^2},\label{eq31}
\end{equation}
which implies
\begin{equation}
\left(\frac{m}{r}\right)_{max} \equiv \frac{M}{R}  \leq \frac{1+ K^2 R^2 -e^{-AR^2}}{2R}.\label{eq32}
\end{equation}
To see the  maximum allowable mass-radius ratio in our model, we
plot  $\frac{m(r)}{r}$ against $r$ (see Fig.~\ref{fig:4}) which
shows that the ratio  $\frac{m(r)}{r}$ is an increasing function
of the radial parameter. We note that a constraint on the maximum
allowed mass-radius ratio in our case falls within the limit   to
the (3+1) dimensional case of  isotropic fluid sphere i.e.,
$\left(\frac{m}{r}\right)_{max} = 0.278 < \frac{4}{9}$.

\subsection{Compactness }
The compactness of the star is given by
\begin{equation}
\label{eq33} u= \frac{ m(r)} {r}=  \frac{1}{2r}\left( 1+K^2 r^2-
e^{-Ar^2}
 \right).
\end{equation}
The surface redshift ($Z_s$) corresponding to the above
compactness ($u$) is obtained as
\begin{equation}
\label{eq34} Z_s= ( 1-2 u)^{-\frac{1}{2}} - 1,
\end{equation}
where
\begin{equation}
\label{eq35} Z_s=  \left[  1-  \frac{1}{r}\left( 1+K^2 r^2-
e^{-Ar^2}
 \right)\right]^{-\frac{1}{2}} - 1 .
\end{equation}
Thus, the maximum surface redshift of our (2+1) dimensional  star
of radius $R = 2.035$ turns out to be $Z_s =  0.51$.

\subsection{Sound speed}
For a physically acceptable model, one expects that the velocity
of sound $v_s=(dp/d\rho)^{\frac{1}{2}}$ should be within the range $0 \leq v_s^2  <1 $. In our
model
\begin{equation}
v_s^2 = \frac{1}{2}
\frac{e^{Ar^2}}{\left(BA+e^{\frac{A}{2}r^2}\right)}\left[
e^{-\frac{A}{2}r^2}+ \frac{1}{\left(BA+e^{\frac{A}{2}r^2}\right)}
\right].\label{eq36}
\end{equation}

\begin{figure}
\begin{center}
\vspace{0.2cm} \includegraphics[width=0.4\textwidth]{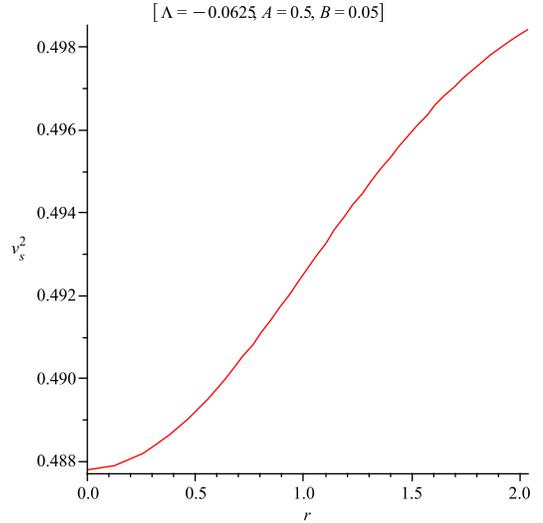}
\end{center}
\caption{The variation of radial sound speed $v_s^2$ is shown
against $r$.}
\label{fig:5}
\end{figure}

We plot the radial sound speed in Fig.~(\ref{fig:5}) and observe
that this parameter satisfies the inequalities $0 < v_s^2 <1$.

\subsection{Equilibrium configuration}
In Section III, we have matched the interior solution to the exterior BTZ metric\cite{BTZ}
across the junction surface $S$ where
 $r=R$. In our case, the junction surface is a one dimensional ring of matter. Though the
 metric coefficients are continuous across
 the surface, their derivatives may not be continuous at the surface. In other words,
  the affine connections may be discontinuous at
 the boundary surface. This can be taken care of if we consider the second fundamental forms of the boundary shell. Let, $\eta$ denotes the Riemann
 normal coordinate at the junction which has positive signature in the manifold
described by exterior BTZ spacetime and negative signature in the
manifold described by the interior spacetime. Mathematically, we
have $x^\mu = ( \tau,\phi,\eta) $  and the normal vector
components $\xi^\mu = ( 0,0,1)$ with the metric

\[  g_{\eta\eta} =1, ~~ g_{\eta i}~ =~0 ~~ and~~ g_{ij} = ~diag~ (~ -1,~r^2
~)~.
\]
\\
 The second fundamental forms  associated with
the two sides of the shell
\cite{Israel1966,Rahaman2006,Rahaman2009,Usmani2010,Rahaman2010b,
Rahaman2011, Perry1992} are then given by
\begin{equation}
K^{i\pm}_j =  \frac{1}{2}\left( g^{ik} \frac{\partial
g_{kj}}{\partial \eta}  \right)_{\eta =\pm 0}
 = \frac{1}{2} \left[\frac{\partial r}{\partial \eta}
 ~ g^{ik} \frac{\partial g_{kj}}{\partial r}\right]_{r=R}.
\label{eq36}
 \end{equation}
The discontinuity in the second fundamental forms is denoted by
\begin{equation}
\kappa _{ij} =   K^+_{ij}-K^-_{ij}.
\label{eq37}
 \end{equation}
Now, from Lanczos equation in (2+1) dimensional spacetime,   one
can obtain the surface stress energy tensor $ S_j^i = diag ( -
\sigma , -v) $      where, $\sigma$ and $v$ are line energy
density and line tension, respectively \cite{Perry1992}

\begin{eqnarray} \sigma &=&  -\frac{1}{8\pi}  \kappa _\phi^\phi,\\
\label{eq38} v &=&  -\frac{1}{8\pi}  \kappa _\tau^\tau.
\label{eq39}
\end{eqnarray}
 Employing Eq.~(\ref{eq36}), we
get
\begin{eqnarray}
\sigma &=&  -\frac{1}{8\pi R}  \left[  \sqrt{ K^2 R^2 - M_0} +  e^{-\frac{1}{2} AR^2}\right],\label{eq40}\\
v &=&  -\frac{1}{8\pi} \left[ \frac{K^2 R} {\sqrt{ K^2 R^2 - M_0}}
+\frac{AR } {BA +e^{\frac{1}{2} AR^2}}\right]\label{eq41}
\end{eqnarray}
where, we have set $r=R$. We note that the line tension is
negative which implies that there must be a line pressure as
opposed to the line tension. As we match the second fundamental
forms, a crucial question arises in the form of the star's
stability against collapse. Therefore, we must have a thin ring of
matter component with above stresses so that the outer boundary
exerts outward force to balance the inward pull of BTZ exterior.
The energy density is negative in this junction ring which is
similar to the $(3+1)$ dimensional case\cite{Visser1989}.

\section{Discussion}
We have obtained a new class of solution corresponding to the BTZ exterior spacetime by assuming a
 particular form of the mass function $m(r)$. The solution obtained is regular at the centre and it
 satisfies all the physical requirements except at the boundary where we propose a thin ring of matter
  content with negative energy density so as to prevent
  collapsing. The discontinuity of the affine connections   at
 the boundary surface provide the above matter confined to the
 ring.
Such a stress-energy tensor is not ruled out from the
consideration Casimir effect for massless fields.
  The solution obtained here has a simple
  analytical form which can be used to study a full collapsing model of a $(2+1)$ dimensional star which is
   beyond the scope of this analysis. This will be taken up else where.

\end{document}